\documentclass[graybox]{svmult}

\usepackage{booktabs}
\usepackage{amsmath}
\usepackage{soul}
\sethlcolor{yellow}
\usepackage{float}
\usepackage{tabularx}
\usepackage{type1cm}        % activate if the above 3 fonts are
\usepackage{makeidx}         % allows index generation
\usepackage{graphicx}        % standard LaTeX graphics tool
\usepackage{multicol}        % used for the two-column index
\usepackage[bottom]{footmisc}% places footnotes at page bottom
\usepackage[numbers,sort&compress]{natbib} % Numeric citations

\usepackage{newtxtext}       % 
\usepackage[varvw]{newtxmath}       % selects Times Roman as basic font
\usepackage[hidelinks,bookmarks=false]{hyperref} \usepackage{xr} 
\hypersetup{%
    bookmarks=false % temporarily moved to usepackage
    pdftitle={Behavioral Prior Elicitation via LLM for Heterogeneous-Agent Bitcoin Flash-Crash Simulation},
    pdfauthor={Pham et al.}%
}

\makeindex             % used for the subject index
\begin{document}

\title*{Behavioral Prior Elicitation via LLM for Heterogeneous-Agent Bitcoin Flash-Crash Simulation}
\titlerunning{LLM-Elicited Priors for Bitcoin Flash-Crash Simulation}

\author{%
Tran Phuong Linh Pham \orcidID{0009-0002-8123-1398} and\\ 
Thi Minh Phuong Ngo \orcidID{0009-0008-2664-1742} and\\ 
Phuong Thao Nguyen \orcidID{0009-0000-3004-9049} and\\ 
Quang Minh Nguyen \orcidID{0009-0006-1135-5861} and\\ 
Thai Hoa Ngo \orcidID{0009-0006-1135-0000} and\\ 
Trong Nghia Nguyen\orcidID{0000-0003-1888-0117}
% \\
% Duc Minh Vu \orcidID{0000-0001-5882-3868}%
}

\authorrunning{Pham et al.}

\institute{Tran Phuong Linh Pham \at National Economics University, Hanoi, Vietnam, \email{11244665@st.neu.edu.vn}
\and Thi Minh Phuong Ngo \at National Economics University, Hanoi, Vietnam, \email{11243142@st.neu.edu.vn}
    \and Phuong Thao Nguyen \at National Economics University, Hanoi, Vietnam, \email{11240203@st.neu.edu.vn}
\and Quang Minh Nguyen \at National Economics University, Hanoi, Vietnam, \email{11247324@st.neu.edu.vn}
\and Thai Hoa Ngo \at National Economics University, Hanoi, Vietnam, \email{hoatn@neu.edu.vn}
\and Trong Nghia Nguyen \at National Economics University, Hanoi, Vietnam, \email{nghiant@neu.edu.vn}
% ,
% \and Duc Minh Vu \at National Economics University, Hanoi, Vietnam,
% \email{minhvd@neu.edu.vn}
}
\maketitle
\abstract{
Bitcoin perpetual futures trade continuously with shallow liquidity and high price impact, making them prone to millisecond-scale flash crashes. Agent-based limit order book simulators are a natural tool for studying such events, but their behavioral parameters are usually hand-crafted or require full order book histories that are rarely available at scale. We propose a four-stage pipeline that constructs a reproducible 100\,ms BTCUSDT flash-crash event panel from 66 detected events, elicits behavioral priors for four agent archetypes using an offline large language model conditioned on empirical microstructure anchors, instantiates a heterogeneous limit order book simulator with the validated priors, and applies causal discovery plus counterfactual analysis to the resulting synthetic panel. Across 500 simulation runs, LLM-elicited priors produce a crash rate of 0.176 versus 0.008 under uninformative uniform priors and generate heavier return tails. In the frozen counterfactual analysis, reducing leverage amplification by one within-sample standard deviation lowers the predicted crash rate from 0.176 to 0.136, whereas reducing directional pressure yields only a marginal change. These findings suggest that empirically anchored LLM elicitation can improve simulator fidelity and help identify plausible crash-amplification mechanisms in a controlled synthetic environment.
}

\section{Introduction}
\label{sec:intro}

Bitcoin perpetual futures occupy a structurally distinct position in modern financial markets because they trade continuously, carry no expiry date, and support persistent leveraged positions, producing order book dynamics that differ from both spot and dated-futures venues. Unlike equity markets supported by a broader and more stable market-making ecosystem~\citep{kirilenko2017}, Bitcoin perpetual futures typically exhibit concentrated liquidity near the best quotes, shallow depth beyond the top levels, and elevated liquidity costs, making them especially vulnerable to flash crashes in which aggressive order flow, liquidity withdrawal, and automated feedback interact to exhaust resting liquidity across multiple price levels. Prior work has identified key flash-crash mechanisms and developed empirical microstructure measures such as Kyle's \(\lambda\)~\citep{kyle1985}, order flow imbalance~\citep{cont2014}, the Amihud illiquidity ratio~\citep{amihud2002}, and VPIN~\citep{easley2012}, but these tools do not directly determine how heterogeneous agents in a simulator should behave. In practice, behavioral inputs are usually supplied either by hand-crafted assumptions inherited from earlier agent-based studies~\citep{lux1999} or by direct calibration from detailed order book data, yet the former lacks market-specific grounding for crypto futures and the latter is often infeasible at the sub-second resolution required for flash-crash analysis. Large language models offer a plausible intermediate route for eliciting behavioral priors, but without empirical discipline their outputs may remain narratively plausible while failing to reproduce the microstructure regularities of the target market. This chapter therefore addresses the need for a framework that combines empirical anchoring, structured LLM-based prior elicitation, heterogeneous-agent simulation, and explicit validation.

This chapter makes four contributions. We construct a reproducible BTCUSDT flash-crash event panel at 100\,ms resolution, spanning 66 identified crash-like events from June 2020 to December 2024 and paired with phase-level microstructure features covering price dynamics, order flow, liquidity, and toxicity proxies (i). We develop a controlled LLM behavioral prior elicitation pipeline in which structured prompts are conditioned on empirical anchors and validated through bounded-support, variance, and cross-archetype sanity constraints before any prior is admitted to the simulator (ii). We instantiate a heterogeneous limit order book simulator with four archetypes: momentum traders, contrarians, HFT market makers, and noise traders, whose parameters are drawn from the validated elicited priors rather than fixed hand-crafted assumptions (iii). We then perform structural mechanism verification and counterfactual sensitivity analysis on the simulated flash-crash panel using DAGMA and DirectLiNGAM, with causal claims restricted to mechanisms operating inside the simulator rather than to real Bitcoin markets (iv).

\section{Related Work}
\label{sec:related}

\textbf{Market microstructure and agent-based modelling.}
Price impact, liquidity withdrawal, and order-flow pressure are central to
flash-crash dynamics. \citet{kyle1985} formalised the \(\lambda\) measure of
price impact, \citet{kirilenko2017} showed that the 2010 flash crash was
amplified by systematic liquidity withdrawal, and \citet{cont2014} established
a strong relation between OFI and price changes. \mbox{\citet{makarov2020}} show that cryptocurrency markets are highly fragmented
across exchanges, with persistent price deviations and strong price sensitivity
to order flow, factors that amplify the risk of rapid dislocations such as flash
crashes. Illiquidity and toxicity are
commonly proxied by \citet{amihud2002} and \citet{easley2012}, respectively.
Agent-based financial models reproduce stylized facts through heterogeneous
traders \citep{lebaron2006}, but their behavioral parameters are typically
hand-crafted and often inherited from equity settings that may not transfer
well to crypto perpetual futures.

\textbf{LLM elicitation and causal discovery.}
Recent work suggests that LLMs can be used as simulated behavioral models under
controlled prompting \citep{horton2023}. We adopt that idea only for offline
prior elicitation, rather than real-time trading, in contrast to autonomous
trading-agent designs such as FinMem \citep{yu2024finmem}. For causal
discovery, NOTEARS and DAGMA provide continuous DAG learning frameworks
\citep{zheng2018,dagma}, while DirectLiNGAM offers a complementary
non-Gaussian identification route \citep{shimizu2011}. To our knowledge, this combination has not previously been applied to an empirically anchored crypto ABM for simulator mechanism verification and counterfactual sensitivity analysis.

\section{Methodology}
\label{sec:method}

\begin{figure}[t]
    \centering
    \Description{Overview of the study pipeline linking empirical flash-crash event extraction, LLM-based behavioral prior elicitation, heterogeneous-agent limit order book simulation, and causal plus counterfactual analysis.}
    \includegraphics[width=\textwidth]{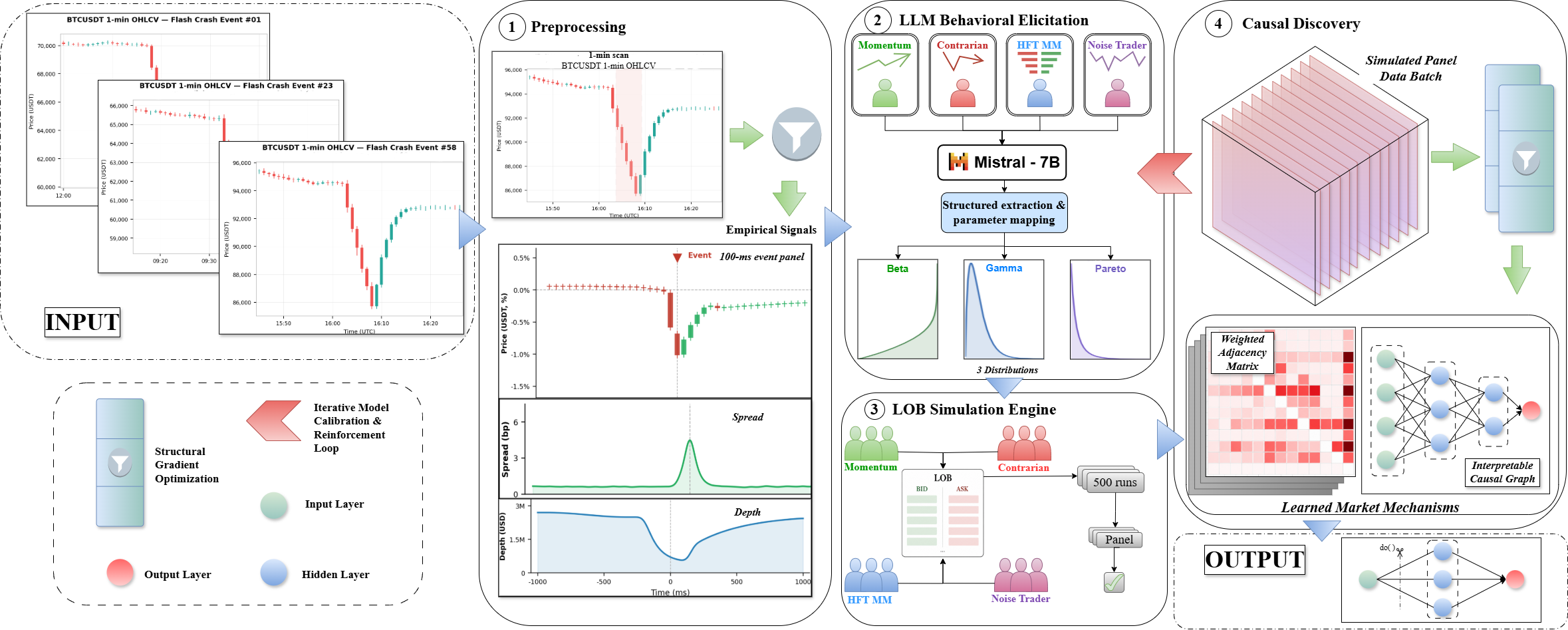}
    \caption{Overall framework of the pipeline. The framework integrates empirical flash-crash event extraction, LLM-conditioned behavioral prior elicitation, heterogeneous-agent LOB simulation, and causal discovery with counterfactual intervention.}
    \label{fig:framework}
\end{figure}

The methodology consists of four stages: empirical event extraction, LLM-based behavioral prior elicitation, heterogeneous-agent LOB simulation, and synthetic-panel causal and counterfactual analysis (Figure~\ref{fig:framework}). 

\subsection{Empirical Flash Crash Event Extraction and Microstructure Characterization}
\label{subsec:empirical}

Raw data are obtained from BTCUSDT perpetual futures via the Tardis API, covering aggregated trades and incremental Level-2 book snapshots at 100\,ms resolution from June 1\textsuperscript{st} 2020 to December 31\textsuperscript{st} 2024. For each event window, we combine a transaction stream, used to identify executed prices, sizes, and aggressor direction, with a quote stream, used to measure spread and best-quote depth. 

\textbf{Flash crash detection.}
We define a flash crash as a transient, partially self-reversing price dislocation: a decline exceeding 3\% within a five-minute window followed by meaningful recovery, consistent with prior flash-crash definitions in market microstructure research~\citep{kirilenko2017}. Let \(P_{\tau}\) denote the transaction price at minute \(\tau\). An event is detected at time \(\tau^*\) if

\[
\frac{P_{\tau^*} - \max_{s \in [\tau^*-5,\,\tau^*]} P_s}
     {\max_{s \in [\tau^*-5,\,\tau^*]} P_s}
\;\leq\; -0.03, \tag{1}
\]

subject to a subsequent retrace of at least 50\% of the realized drawdown within the recovery window. Overlapping detections within four hours are deduplicated by retaining the largest event in each cluster. This procedure yields an event corpus \(\mathcal{E}\) of 66 identified crash-like BTCUSDT events, of which 43 satisfy the full retrace criterion and 23 are steep trend-drop entries retained for behavioral baseline coverage.  For each \(e \in \mathcal{E}\), we reconstruct a symmetric \(\pm 30\)-minute window around the event and retain a separate non-crash reference set for baseline calibration.

\textbf{Event phase labeling.}
Each event is partitioned into four ordered phases,
\(\Phi=\{\textit{pre},\textit{drop},\textit{recovery},\textit{post}\}\), so that downstream aggregation follows crash anatomy rather than calendar time. The crash onset \(t_0\) is the local peak in a five-minute lookback window, and \(t_{\min}\) is the subsequent trough. Writing \(P_0=P_{t_0}\) and \(P_{\min}=P_{t_{\min}}\), the recovery boundary \(t_r\) is the first \(t>t_{\min}\) such that

\[
\frac{P_t - P_{\min}}{P_0 - P_{\min}}
\;\geq\; 0.20
\quad \text{for at least 3 consecutive 100\,ms bins.} \tag{2}
\]

The pre-phase spans the 60 seconds preceding \(t_0\), and the post-phase runs from \(t_r\) to the end of the observation window. These phase labels are used both for empirical aggregation and for simulation initialization.

\textbf{Exogenous catalyst classification.}
Each event is assigned a binary news indicator \(D_{\text{news}}\) based on a curated set of BTC-relevant regulatory, exchange, and macro events within a \(\pm 1\)-day window. Because timestamp-level causal ordering is unavailable, this variable is used descriptively rather than for causal claims.

\textbf{Microstructure feature construction.}
All features are computed at 100\,ms resolution to match the simulation clock and the downstream analysis. The primary directional-pressure measure is order-flow imbalance,

\[
\mathrm{OFI}_t
\;=\;
\sum_{i \in \mathrm{bin}_t} \mathrm{sgn}_i\, q_i, \tag{3}
\]

where \(\mathrm{sgn}_i \in \{+1,-1\}\) denotes aggressor direction and \(q_i\) is traded quantity. We also track spread, depth imbalance, volatility, trade intensity, and a rolling Kyle-style price-impact coefficient~\citep{kyle1985},

\[
\lambda_t
\;=\;
\frac{\mathrm{Cov}\!\left(\Delta P_t,\mathrm{OFI}_t\right)}
     {\mathrm{Var}\!\left(\mathrm{OFI}_t\right)}. \tag{4}
\]

The resulting representation \(\mathbf{X}_{\mathrm{emp}}\) contains phase-labeled 100\,ms state vectors, event boundaries, and \(D_{\text{news}}\). Phase-level summaries \(\boldsymbol{\mu}_{\phi}=\mathbb{E}_{\phi}[\mathbf{x}_t]\) are used to condition elicitation.

\subsection{LLM-Conditioned Behavioral Prior Elicitation}
\label{subsec:elicitation}

The phase-level empirical summaries \(\boldsymbol{\mu}_{\phi}\) describe how BTC flash-crash microstructure behaves in each phase, but they do not specify how heterogeneous agents should act under those conditions. We use an offline LLM (Mistral-7B) to elicit phase-conditional behavioral priors from sampled empirical states and phase summaries \(\boldsymbol{\mu}_{\phi}\). Responses are parsed into structured parameter proposals and used only for prior construction, following recent work on controlled LLM-based behavioral modeling~\citep{horton2023}.

\textbf{Archetypes, prompt design, and validation.}
We consider four trader archetypes: momentum traders, noise traders, market makers, and contrarian traders. For each archetype \(a \in \mathcal{A}\) and phase \(\phi \in \Phi\), the elicitation targets parameters such as aggressiveness, cancel probability, inventory sensitivity, and order size.  Market makers additionally receive a volatility-withdrawal threshold \(\sigma_a^*\), while the noise-trader arrival rate is assigned directly from phase-level empirical trade intensity because it is treated as a market-level observable rather than a latent behavioral trait. Prompts are conditioned on sampled empirical states and phase summaries, including OFI quantiles, realized volatility, Kyle's \(\lambda\), and bid-ask spread. Responses that violate predefined validity rules are removed before fitting. 

\textbf{Behavioral prior construction.}
Valid responses are fit by maximum likelihood to parametric families appropriate for bounded, positive, and heavy-tail variables, yielding one fitted prior per archetype-phase pair:
\begin{equation}\tag{5}
    \boldsymbol{\pi}_a(\phi)
    \;=\;
    \bigl\{\, p(\theta_j \mid a,\phi) \bigr\}_{j},
    \label{eq:prior}
\end{equation}
with one fitted marginal distribution per behavioral parameter. After fitting, the priors are frozen and no further LLM interaction occurs in the pipeline. These priors are the only LLM-derived inputs passed to the simulator. Elicitation was performed with a single model at a fixed temperature, so cross-model robustness and temperature sensitivity remain future work.

\subsection{Heterogeneous-Agent Limit Order Book Simulation}
\label{subsec:simulation}

The priors \(\boldsymbol{\pi}_a(\phi)\) specify phase-conditional behavior, but they do not by themselves generate order flow or price dynamics. We therefore instantiate them in a discrete-time, two-sided limit order book with heterogeneous agents, run \(R=500\) independent realizations, and record the resulting synthetic panel for downstream analysis.

\textbf{Agent setup.}
Each run contains \(N=100\) agents drawn from four archetypes: momentum traders, noise traders, market makers, and contrarian traders. At initialization, every agent \(i\) of archetype \(a(i)\) draws a static parameter vector
\begin{equation}\tag{6}
    \boldsymbol{\theta}_i \;\sim\; \boldsymbol{\pi}_{a(i)}(\phi),
    \label{eq:sample}
\end{equation}
with pre-crash priors used by default and phase-specific priors activated as the run progresses. Static within-run sampling is deliberate because the downstream causal analysis treats each realization as a stable data-generating system. 

\textbf{Order generation and price update.}
The exchange operates at a 100\,ms tick interval, matching the empirical panel. At each tick \(t\), agent activation follows a Poisson process calibrated to phase-level trade intensity. Active agents choose order type, direction, and size according to their archetype-specific rules and their sampled parameter vector \(\boldsymbol{\theta}_i\): momentum agents trade with the trend, noise agents trade randomly, market makers post passive quotes, and contrarian agents buy against sufficiently deep drawdowns. Price impact follows the linearized formulation of \citep{kyle1985}. Let \(\mathrm{NF}_t\) denote signed net order flow, \(\lambda_{\phi}\) the phase-level impact coefficient inherited from \(\boldsymbol{\mu}_{\phi}\), \(\kappa_{\phi}\) a phase-specific impact scale, and \(\rho_t \in (0,1]\) a resilience term. The mid-price update is
\begin{equation}\tag{7}
    \Delta P_t \;=\; \lambda_{\phi} \cdot \mathrm{NF}_t \cdot \kappa_{\phi} \cdot \rho_t.
    \label{eq:price}
\end{equation}
As resilience weakens, the same amount of net order flow produces a larger price change.

\textbf{Liquidity withdrawal, recovery, and run initialization.}
Market makers are the main amplification channel in the simulator. Each market maker \(i\) draws a volatility-withdrawal threshold \(\sigma_i^*\); if realized volatility over the preceding window exceeds that threshold, the agent cancels outstanding quotes and stops providing liquidity. As volatility rises during the drop phase, market makers withdraw sequentially, widening spreads and thinning depth. Recovery is driven by contrarian agents, which accumulate long inventory against the falling price. Each of the \(R=500\) runs is initialized from a randomly selected event \(e \in \mathcal{E}\), inheriting that event's observed phase durations and phase-level summaries as initial targets. Cross-run variation therefore comes from stochastic event selection and behavioral draws, not from changing the underlying market definition. 

\textbf{Synthetic panel construction.}
At each tick \(t\) of each censored run \(r\), the simulator records
\[
    \mathbf{s}_{r,t}
    \;=\;
    \bigl(
        \mathrm{OFI}_t,\; s_t,\; d_t,\;
        \Lambda_t,\; \mathrm{VPIN}_t,\; W_t,\;
        \mathbb{1}_{\mathrm{crash}}
    \bigr),
\]
where \(s_t\) denotes bid-ask spread, \(d_t\) depth imbalance, \(\Lambda_t = |\dot{P}_t| \cdot |\ddot{P}_t| / \hat{\sigma}_t\) is the leverage proxy following \citep{brunnermeier2009}, \(W_t\) is aggregate agent wealth, and \(\mathbb{1}_{\mathrm{crash}}\) is the binary flash-crash indicator. In the absence of direct book observations for every feature, \(d_t\) is constructed from trade-flow data as a normalized function of \(\mathrm{OFI}_t\), so any recovered edge from depth imbalance to the crash indicator should be interpreted as partly overlapping the OFI pathway rather than as a fully independent liquidity channel.

For the primary causal analysis, each run is compressed to a single summary row using the same fair window for crash and non-crash realizations: the last 200 pre-phase observations, ending 20 ticks before the crash detector in crash runs and 20 ticks before the end of the pre phase in non-crash runs. Mean summaries are used for directional pressure, book imbalance, mean wealth, and wealth concentration, while the leverage proxy enters through its pre-crash maximum. Candidate summaries with negligible cross-run variance are removed before learning. The resulting run-level panel \(\mathbf{Z} \in \mathbb{R}^{R \times p}\) is the input to the graph-learning stage.

\subsection{Structural Mechanism Verification on the Synthetic Simulation Panel}
\label{subsec:causal}

% We apply graph learning to the run-level panel \(\mathbf{Z}\) to test whether standard methods recover the main simulator-embedded crash pathways from cross-run variation. This is a simulator-internal mechanism check rather than a claim about real-market causality. Before estimation, continuous variables are standardized and near-constant columns are removed.

The run-level panel \(\mathbf{Z} \in \mathbb{R}^{R \times p}\) summarizes, for each of the \(R=500\) simulated realizations, the pre-crash state preceding crash onset or the matched terminal window in non-crashing runs. This stage tests whether standard graph-learning methods recover the main pathways embedded in the simulator, not causal structure in the real market.

This design is necessary because historical BTC market data does not provide the controlled variation needed for credible causal recovery: order flow, leverage, inventories, and external shocks co-vary in ways that cannot be disentangled observationally. By contrast, the synthetic panel is generated by a known mechanism in which directional pressure, liquidity withdrawal, spread widening, depth deterioration, and crash occurrence are linked by construction. Recovering those links from \(\mathbf{Z}\) therefore serves as a simulator-internal mechanism check rather than as a claim about real-market causality.

Before graph learning, each continuous column of \(\mathbf{Z}\) is standardized to zero mean and unit variance, and near-constant columns are removed for numerical stability. We work with the run-level cross section because the inferential target is whether pre-crash directional pressure and leverage amplification are associated with crash occurrence across independent realizations. 

Two complementary graph-learning methods are applied to the standardized panel. The first is DAGMA~\citep{dagma}, a continuous acyclicity-constrained estimator in the NOTEARS family~\citep{zheng2018}. The estimator solves
\begin{equation}\tag{8}
    \min_{\hat{W}} \;
    \frac{1}{2R}\bigl\|\mathbf{Z} - \mathbf{Z}\hat{W}\bigr\|_F^2
    \;+\; \lambda_1 \|\hat{W}\|_1
    \quad \text{s.t.} \quad
    h(\hat{W}) = 0,
    \label{eq:dagma}
\end{equation}
where \(\|\cdot\|_F\) denotes the Frobenius norm, \(\lambda_1\) controls edge sparsity, and \(h(\hat{W})=0\) enforces acyclicity~\citep{dagma}. The second is DirectLiNGAM~\citep{shimizu2011}, which provides an independent orientation check based on non-Gaussian residual structure under a linear acyclic model. 

Both methods yield directed graphs over the retained variables of \(\mathbf{Z}\). The structural representation \(\hat{\mathcal{G}}\) passed to Sect.~\ref{subsec:intervention} is the graph recovered by the primary continuous-optimization approach, while the DirectLiNGAM graph is used as a directionality cross-check. In the final paper, we use the two methods comparatively rather than score them against a hand-coded ground-truth DAG. The main robustness criteria are whether crash-adjacent edges survive the fair-window design, whether the two methods agree on the outcome-adjacent structure, and whether the descriptive separation results in Sect.~\ref{subsec:res-dag} support the same qualitative mechanism.

\subsection{Counterfactual Intervention Analysis}
\label{subsec:intervention}

Using \(\hat{\mathcal{G}}\) and the run-level panel \(\mathbf{Z}\), we estimate local counterfactual sensitivity of crash risk to reductions in the main pre-crash channels within the observed support.

We consider two mitigation targets: \(\overline{\mathrm{OFI}}\), the mean directional-pressure summary, and \(\Lambda_{\max}\), the peak leverage proxy. A logistic regression model is fitted on \(\mathbf{Z}\) with binary crash outcome \(y_r \in \{0,1\}\). Let \(\mathbf{x}_r \in \mathbb{R}^{p}\) denote the retained feature vector for run \(r\). In the main specification, the feature set includes directional pressure, book imbalance, leverage, mean wealth, and wealth concentration, with all features standardized before estimation. The model is
\begin{equation}\tag{9}
    \hat{P}(y_r = 1 \mid \mathbf{x}_r)
    \;=\;
    \sigma\!\left(\mathbf{x}_r^\top \hat{\boldsymbol{\beta}}\right),
    \label{eq:logit}
\end{equation}
where \(\sigma(\cdot)\) is the logistic function and \(\hat{\boldsymbol{\beta}}\) is estimated by maximum likelihood. 

For each target feature \(x_{rj}\), the counterfactual assignment reduces the observed value by one within-panel standard deviation while preventing it from falling below the non-crash median:
\begin{equation}\tag{10}
    \tilde{x}_{rj}
    \;=\;
    \max\!\left\{x_{rj} - \sigma_j,\; m^{(0)}_j\right\},
    \label{eq:counterfactual}
\end{equation}
where \(\sigma_j\) is the within-panel standard deviation of feature \(j\) and \(m^{(0)}_j\) is the corresponding non-crash median. The counterfactual crash probability is then
\begin{equation}\tag{11}
    \hat{P}_{\mathrm{cf}}(y_r = 1)
    \;=\;
    \sigma\!\left(\tilde{\mathbf{x}}_r^\top \hat{\boldsymbol{\beta}}\right).
\end{equation}
These are local regression-based sensitivity calculations on the frozen run panel, not full structural interventions propagated through every edge of \(\hat{\mathcal{G}}\).

The population-level counterfactual crash rate is the mean of \(\hat{P}_{\mathrm{cf}}(y_r = 1)\) across runs. Let
\(\bar{P}_{\mathrm{obs}} = \frac{1}{R}\sum_r \hat{P}(y_r=1 \mid \mathbf{x}_r)\)
denote the mean model-predicted crash rate under the observed panel. The relative crash-rate reduction under mitigation of feature \(j\) is
\begin{equation}\tag{12}
    \hat{Y}_{\mathrm{mitigate}}^{\,j}
    \;=\;
    \frac{\bar{P}_{\mathrm{obs}} - \bar{P}_{\mathrm{cf}}^{\,j}}{\bar{P}_{\mathrm{obs}}},
    \label{eq:effect}
\end{equation}
for \(j \in \{\overline{\mathrm{OFI}}, \Lambda_{\max}\}\). In the selected frozen specification, leverage mitigation is interpreted as a stronger local effect than OFI mitigation, but not as evidence that removing a single predictor would eliminate crashes. 

% ============================================================
% 4. EXPERIMENTS
% ============================================================
\section{Experiments}
\label{sec:exp}

This section defines the frozen evaluation protocol used throughout the paper, including the locked BTCUSDT panel, the comparison baselines, and the run-level design used for mechanism and intervention analysis.

\subsection{Data, Baselines, and Freeze Protocol}
\label{subsec:data}
\label{subsec:baselines}

We evaluate the simulator on a locked BTCUSDT flash-crash panel containing 66 events aligned on a 100\,ms grid, and all reported results use the same authoritative 500-run simulation freeze. The empirical references are heterogeneous by design: return tails and phase-level directional flow are taken from the crash-conditioned event panel, whereas crash incidence is evaluated against a target band rather than a single empirical point estimate. The primary condition uses offline LLM-elicited behavioral priors frozen before simulation and is compared against two locked baselines, a broad random-prior condition and a literature-prior condition based on hand-crafted values from prior simulation studies. All three conditions share the same event panel, 100\,ms clock, and seed discipline. We regard the LLM condition as successful if it dominates the broad-prior baseline and remains non-inferior to the literature baseline on crash incidence and mean short-horizon drawdown, while preserving the correct sign ordering of directional flow between the pre-crash and drop phases. 

\subsection{Evaluation Metrics}
\label{subsec:metrics}

The frozen evaluation stack covers five layers: distributional realism, phase fidelity, crash realism, mechanism evidence, and counterfactual effect size. 

\subsection{Run-Level Design}
\label{subsec:runlevel}

A run is labeled as a crash if price falls by at least 1.93\% within a 10-step window, corresponding to one second on the 100\,ms grid. This threshold is calibrated from the empirical event distribution, where it matches the 10th percentile of the rolling-10-tick drawdown distribution across the 66-event panel. It is distinct from the empirical event detector in Sect.~\ref{subsec:empirical} (\(\geq\)3.0\% drawdown within 5 minutes with \(\geq\)50\% recovery): the empirical detector selects historical episodes for the catalog, whereas the simulated detector only classifies individual simulation realizations.

Both crash and non-crash runs are summarized using the same trailing window: the last 200 pre-phase observations, ending 20 ticks before the crash detector in crash runs and 20 ticks before the end of the pre phase in non-crash runs. Under this design, the locked LLM panel contains 500 run-level observations, of which 88 are crash runs and 412 are non-crash runs. The run-level panel is used for two purposes: structure recovery with two complementary algorithms, and descriptive separation plus single-feature discrimination analysis for the main pre-crash summaries. Mitigation follows the support-preserving construction in Eq.\ref{eq:counterfactual}

% ============================================================
% 5. RESULTS
% ============================================================
\section{Results}
\label{sec:results}

We report the frozen comparison in three layers: stylized-fact and crash-dynamics realism, run-level mechanism recovery, and counterfactual intervention effects.

\subsection{Stylized-Fact and Flash-Crash Realism}
\label{subsec:res-stylized}

Table~\ref{tab:locked-comparison} summarizes the locked three-way comparison. The LLM condition clearly improves over the broad-prior baseline on crash incidence and mean short-horizon drawdown, and it remains competitive with the literature baseline on those same crash-level metrics. It does not, however, reproduce the full tail magnitude of the crash-conditioned BTC reference panel and it overshoots the empirical directional-flow magnitude during the drop phase.

\begin{table*}
\centering
\caption{Locked three-condition comparison on the frozen 500-run panel.}
\label{tab:locked-comparison}
{\scriptsize
\setlength{\tabcolsep}{4pt}
\begin{tabularx}{\textwidth}{@{}>{\raggedright\arraybackslash}p{3.0cm}>{\centering\arraybackslash}p{1.6cm}>{\centering\arraybackslash}p{1.0cm}>{\raggedright\arraybackslash}X>{\centering\arraybackslash}p{1.1cm}>{\centering\arraybackslash}p{1.4cm}@{}}
\specialrule{0.08em}{0pt}{0pt}
{\bfseries Criterion} & \textbf{Empirical ref.} & \textbf{LLM} & \textbf{Gap (LLM vs. ref.)} & \textbf{Broad} & \textbf{Literature} \\
\midrule
Tail thickness of returns & 8581.7880 & 29.0842 & 99.66\% below empirical; about 295$\times$ smaller & 26.4920 & 77.1955 \\
Short-run volatility persistence & 1.0000 & 1.0000 & matched & 1.0000 & 1.0000 \\
Directional flow, pre-crash phase & 0.2644 & 0.1869 & 29.32\% below empirical & 0.1154 & 0.1848 \\
Directional flow, drop phase & -2.6988 & -5.0391 & 86.71\% more negative than empirical & -2.9120 & -2.8810 \\
Crash incidence & 0.05 to 0.40 band & 0.1760 & n/a & 0.0080 & 0.1520 \\
Mean short-horizon drawdown (\%) & n/a & 1.3280 & n/a & 0.6122 & 1.1276 \\
\bottomrule
\end{tabularx}
}
\end{table*}

The tail reference should be read as an upper-envelope benchmark for extreme flash-crash behavior rather than as a point target, because it is computed from a crash-conditioned 100\,ms event panel. Overall, the evidence supports a measured claim: empirically disciplined LLM priors improve fidelity relative to an undisciplined broad baseline, while remaining competitive rather than uniformly dominant relative to the literature baseline.

\begin{figure}[t]
  \centering
  \Description{Temporal evolution of price, directional pressure, quoted trading cost, book balance, and trading activity around the first detected crash in the locked LLM panel.}
    \IfFileExists{figures/paper/figure_02_temporal_anatomy_llm.pdf}{%
        \includegraphics[width=0.8\linewidth]{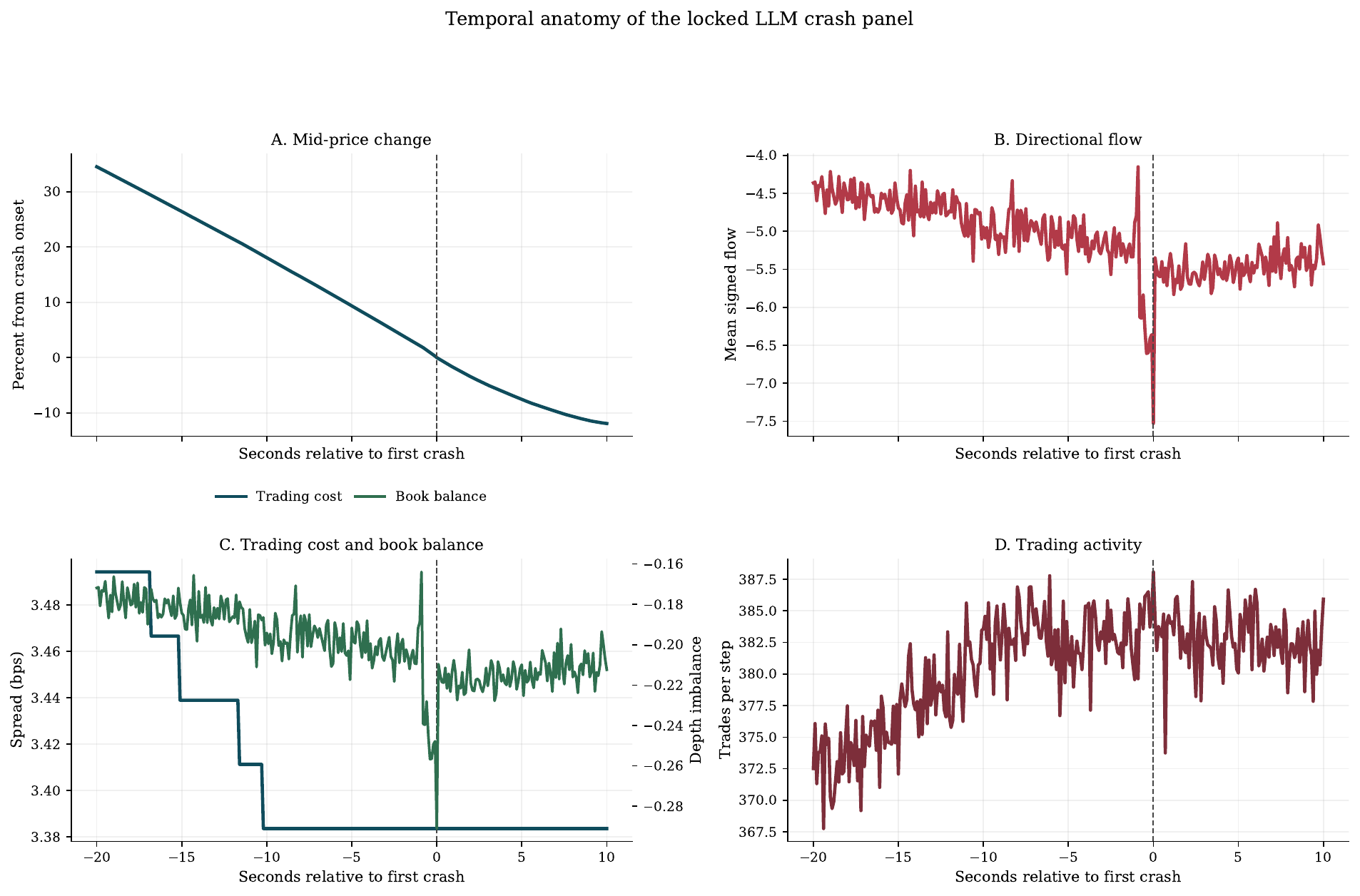}%
    }{%
        \fbox{\parbox{0.8\linewidth}{\centering Placeholder for missing figure:\\\texttt{\detokenize{figures/paper/figure_02_temporal_anatomy_llm.pdf}}}}%
    }
  \caption{Temporal anatomy of the locked LLM crash panel after alignment at the first detected crash. Price, directional pressure, quoted trading cost, book balance, and trading activity co-move around the event window.}
  \label{fig:temporal-anatomy}
\end{figure}

Figure~\ref{fig:temporal-anatomy} shows that the locked LLM panel generates crash events through a coordinated sequence rather than isolated price jumps. Directional pressure turns sharply negative before and during the drop, quoted trading cost rises as book balance deteriorates, trading activity accelerates into the event window, and price then falls in a concentrated burst with only partial early recovery. This should be interpreted as correct sequencing and co-movement, not perfect amplitude calibration.

\subsection{Structural Consistency Verification}
\label{subsec:res-dag}

The run-level analysis provides partial rather than complete recovery of the hypothesized mechanism, but the surviving signal is more credible under the symmetric window construction. In the main specification, DAGMA retains a direct edge from pre-crash leverage into the crash label, while DirectLiNGAM assigns the most stable crash-adjacent role to wealth-level summaries. Directional pressure and book imbalance remain strongly separated across crash and non-crash runs, but they do not recover as uniformly oriented crash edges once the panel is no longer label-conditioned by construction. Quoted trading cost and Kyle-style impact become effectively constant inside the strict pre-phase window and are therefore removed by the variance filter.

Table~\ref{tab:runlevel-stats} provides descriptive support for this result. Leverage rises from 1.1115 in non-crash runs to 1.3446 in crash runs, wealth concentration falls from 0.9456 to 0.9128, and mean wealth rises modestly from 175485.82 to 179189.33. Mann-Whitney tests remain overwhelmingly significant for the main summaries, while the strongest single-feature discrimination scores now top out around 0.98 rather than 1.00. This is consistent with the intended interpretation: the panel retains strong pre-crash signal without becoming perfectly separable by construction.

\begin{table*}
\centering
\small
\caption{Run-level descriptive support for the mechanism analysis on the locked LLM panel.}
\label{tab:runlevel-stats}
\begin{tabular*}{\textwidth}{@{\extracolsep{\fill}}lcccc@{}}
\toprule
\textbf{Pre-crash summary} & \textbf{Non-crash mean} & \textbf{Crash mean} & \textbf{Mann-Whitney $p$} & \textbf{Best AUC} \\
\midrule
Directional pressure & 0.0112 & 0.2646 & 4.07e-39 & 0.944 \\
Book imbalance & 0.0004 & 0.0106 & 4.15e-39 & 0.944 \\
Leverage peak & 1.1115 & 1.3446 & 6.11e-46 & 0.983 \\
Mean wealth & 175485.82 & 179189.33 & 1.77e-45 & 0.980 \\
Wealth concentration & 0.9456 & 0.9128 & 2.39e-45 & 0.980 \\
\bottomrule
\end{tabular*}
\end{table*}

The most defensible conclusion is partial structural consistency. The structure-learning methods recover a subset of the embedded crash pathways from cross-run variation, with leverage emerging as the most stable simulator-internal amplifier signal. The full cascade from flow pressure to trading-cost widening to book thinning to crash realization is not recovered in a uniformly stable way.

\subsection{Counterfactual Results}
\label{subsec:ablation}

\begin{figure}[H]
  \centering
    \Description{Pre-crash separation and local mitigation effects on the symmetric trailing-window LLM panel, showing directional pressure as a warning signal and leverage as the cleaner local amplifier.}
    \IfFileExists{figures/paper/figure_03_precrash_and_interventions.pdf}{%
        \includegraphics[width=0.98\linewidth]{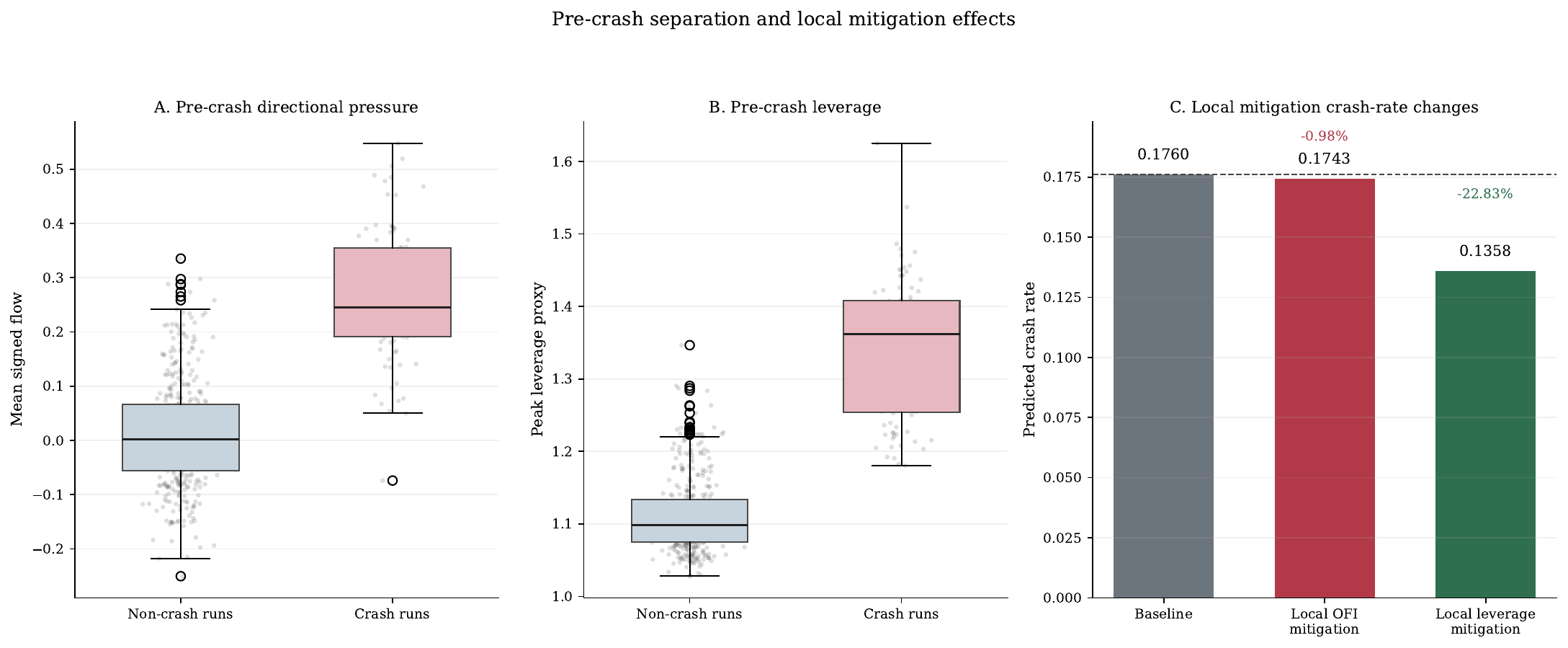}%
    }{%
        \fbox{\parbox{0.94\linewidth}{\centering Placeholder for missing figure:\\\texttt{\detokenize{figures/paper/figure_03_precrash_and_interventions.pdf}}}}%
    }
  \caption{Pre-crash separation and local mitigation effects on the symmetric trailing-window LLM panel. Directional pressure remains a strong warning signal, but leverage is the cleaner local amplifier once the intervention is kept on-manifold.}
  \label{fig:precrash-intervention}
\end{figure}

We report only local, support-preserving interventions in the main text, because earlier zero-out interventions move predictors too far outside the support of the frozen run panel. Table~\ref{tab:interventions} shows that mitigating directional pressure changes the predicted crash rate only from 0.1760 to 0.1743, a reduction of 0.98\%, whereas mitigating leverage lowers it to 0.1358, a reduction of 22.83\%. Across the 0, 20, and 50 tick gap ablations with the broader feature set, the leverage effect remains in the 20 to 23\% range while the directional-pressure effect remains close to zero. This contrast is also visible in Fig.~\ref{fig:precrash-intervention}, which shows strong pre-crash separation for directional pressure but a cleaner local mitigation effect for leverage.

\begin{table}[H]
\centering
\caption{Counterfactual intervention results on the locked run-level suite.}
\label{tab:interventions}
\begin{tabular}{lccc}
\toprule
\textbf{Intervention} & \textbf{Baseline rate} & \textbf{Post-change rate} & \textbf{Change} \\
\midrule
Local mitigation of directional pressure & 0.1760 & 0.1743 & -0.98\% \\
Local mitigation of leverage amplification & 0.1760 & 0.1358 & -22.83\% \\
\bottomrule
\end{tabular}
\end{table}

\section{Discussion}
\label{sec:discuss}

\begin{sloppypar}
The central finding of this chapter is not that LLM outputs are behaviorally true, but that LLM-elicited behavioral priors become useful simulator inputs when they are constrained by empirical anchors and evaluated against explicit baselines. In the frozen comparison, the LLM condition attains a crash rate of~0.176, well inside the target band of~0.05--0.40, whereas the broad uniform baseline produces only~0.008 and a substantially smaller mean short-horizon drawdown (0.612 vs.~1.328), indicating that undisciplined priors do not reproduce realistic crash dynamics. Relative to the literature baseline, however, the comparison is more balanced: literature priors achieve competitive crash incidence and drawdown while generating much heavier tails (excess kurtosis 77.20 vs.~29.08), so the appropriate interpretation is competitiveness rather than uniform superiority. The persistent gap in tail thickness warrants mechanistic explanation. Four structural factors account for most of the 295-fold shortfall. First, our price-update rule Eq.\mbox{\ref{eq:price}} is linear in net order flow; because Poisson-activated order flow  aggregates near-Gaussian across 100 agents by the central limit theorem, a  linear map largely preserves distributional thinness,  whereas empirically documented market impact is concave in trade size and is itself a  recognised amplifier of tail thickness~\mbox{\citep{bouchaud2004nonlinear,gabaix2003theory}}. Second, static within-run parameter sampling Eq.\ref{eq:sample}  forecloses real-time strategy switching and herding contagion, a channel that the agent-based literature identifies as the primary endogenous source of excess kurtosis~\citep{lux1999,cont2000herd}. Third, because archetypes act independently, the model limits the probability of large-scale synchronised liquidity withdrawal, precisely the configuration responsible for the most extreme empirical moves ~\citep{kirilenko2017,lebaron2006}. Fourth, the sanity-check gates applied to LLM responses (Sect.~\ref{subsec:elicitation}) remove rare, extreme behavioral proposals in favour of well-behaved bounded-support priors, trading tail fidelity for elicitation reliability. An additional conditioning asymmetry inflates the gap: the empirical kurtosis is computed from a crash-conditioned 100\,ms panel and should be read as an upper-envelope benchmark, whereas simulated kurtosis pools crash and non-crash runs alike. Closing this gap through non-linear impact specifications, dynamic strategy switching, and explicit inter-agent contagion represents the most tractable direction for future simulator extensions. A naturalkirilenko2017,lebaron concern is that LLM elicitation may generate parameters that are narratively plausible but quantitatively invalid, yet in this framework that risk is reduced by a constrained JSON schema, bounded-support distributions, pre-simulation sanity checks, and frozen prompt outputs; the ablation evidence is consistent with that design, because the elicited priors improve clearly over the broad baseline without uniformly dominating informed hand-crafted alternatives. At the same time, the conclusions remain bounded by four limitations: several variables are fallback constructions from trade-level data rather than canonical Level-2 measurements, so measurement error can propagate into both the empirical anchors and the simulated panel; two planned robustness checks, namely a no-phase-conditioning ablation and a uniform-prior causal suite, are not yet available in authoritative frozen form; and several important drivers of crypto flash crashes, including structured news or sentiment inputs, direct liquidation feeds, and canonical Level-2 order book reconstructions, remain absent from the current pipeline. Relatedly, the fixed population of 100 agents across four archetypes is a simplification relative to the scale and behavioral diversity of actual BTC perpetual futures participants. Accordingly, all causal claims in this chapter are restricted to mechanisms operating within the simulator and should not be interpreted as causal claims about real Bitcoin markets.
\end{sloppypar}

\section{Conclusion}
\label{sec:conclusion}

\begin{sloppypar}
This capter presents a simulation-first framework that transforms empirical flash-crash evidence into behaviorally grounded agent priors, validates those priors against microstructure anchors, instantiates them in a heterogeneous limit order book simulator, and studies crash-amplification mechanisms through structural verification and local counterfactual sensitivity analysis on the resulting synthetic panel. Its main contributions are a 66-event BTCUSDT flash-crash panel at 100\,ms resolution,  a structured offline LLM elicitation pipeline with sanity gates and ablation baselines, a four-archetype heterogeneous LOB simulator parameterised by validated empirical anchors, and a run-level verification suite combining DAGMA and DirectLiNGAM with regression-based local mitigation analysis. Across the authoritative 500-run canonical batch, LLM-elicited priors substantially outperform broad uniform priors and remain competitive with literature priors on crash realism, with leverage mitigation producing a markedly stronger local crash-rate reduction than directional-pressure mitigation. Within the simulator, leverage amplification therefore emerges as the more consequential local crash-amplification channel. All causal claims are restricted to the simulated environment.
\end{sloppypar}

\ethics{Ethics Approval}{This chapter uses exchange market data only and does not involve human participants, human tissue, or animals. Ethics approval and informed consent were therefore not required.}

\end{document}